\documentclass[conference]{IEEEtran}
\newcommand{\trtitle}{Reporting and Analysing the Environmental Impact of Language Models on the Example of Commonsense Question Answering with External Knowledge}
\newcommand{\titleshort}{Reporting and Analysing the Environmental Impact of Language Models}

\newcommand{\trgroup}{Semantic Systems Research Group, SEMS}
\newcommand{\truniversity}{Department of Informatics, University of Hamburg}
\usepackage[main=english]{babel}
\usepackage{csquotes}
\usepackage{placeins}
\usepackage{float}
\usepackage{booktabs,tabularx}
\usepackage{siunitx}
\usepackage{etoolbox}
\usepackage{adjustbox}
\usepackage{svg}
\usepackage{lipsum}
\usepackage{mathrsfs}
\usepackage{acronym}                    
\usepackage{algorithmic}			    
\usepackage{algorithm}					
\usepackage{amsfonts}                   
\usepackage{amsmath}                    
\usepackage{amssymb}                    
\usepackage{amsthm}
\usepackage{xcolor}                     
\usepackage{booktabs}                   
\usepackage{threeparttable}             
\usepackage{nicematrix}                 
\usepackage{tikz}                       
\usepackage{color}                      
\definecolor{uhhRed}{RGB}{254,0,0}		
\definecolor{uhhGrey}{RGB}{122,122,120} 
\definecolor{l8Grey}{RGB}{232,232,230}  
\usepackage{fancybox}                   
\usepackage{fancyhdr}				    
\usepackage{geometry} 	                
\usepackage{longtable}					
\usepackage{listings}                   
\usepackage{multicol}					
\usepackage{multirow}					
\usepackage{rotating}					
\usepackage{silence}                    
\usepackage{subcaption}                 
\usepackage{tabularx}					
\usepackage{url,xspace,boxedminipage}   
\usepackage[style=ieee,citestyle=numeric-comp,sorting=ynt]{biblatex}
\DeclareGraphicsExtensions{.pdf,.svg,.jpg,.png,.eps} 
\graphicspath{{./src/}} 				
\providecommand{\keywords}[1]
{
  \small	
  \textbf{\textit{Keywords---}} #1
}
\theoremstyle{plain}

\theoremstyle{definition}

\theoremstyle{remark}

\newcolumntype{x}[1]{>{\centering\arraybackslash}p{#1}}     
\usepackage{graphicx}                                       
\usepackage{hyperref}                                       
\hypersetup{
    colorlinks=true,
    linkcolor=blue,
    filecolor=blue,      
    urlcolor=blue,
    citecolor=blue
    }
\makeatletter
\def\footnoterule{\kern-3\p@ \hrule \kern2.6\p@}            
\makeatother
\makeatletter
\newcommand{\linebreakand}{
  \end{@IEEEauthorhalign}
  \hfill\mbox{}\par
  \mbox{}\hfill\begin{@IEEEauthorhalign}
}
\makeatother
\begin{document}
\title{\trtitle}
\renewcommand{\headheight}{28pt}
\fancyhead[LO]{\slshape}
\fancyhead[RO]{ \slshape \titleshort}
\author{Aida Usmanova, Junbo Huang, Debayan Banerjee, Ricardo Usbeck
\linebreakand
\trgroup
\linebreakand
\truniversity
\linebreakand
\texttt{aida.usmanova@studium.uni-hamburg.de,}
\linebreakand
\texttt{(junbo.huang, debayan.banerjee, ricardo.usbeck)@uni-hamburg.de}
}
\maketitle
\hypertarget{Abstract}{}
\begin{abstract}
\label{Abstract}
Human-produced emissions are growing at an alarming rate, causing already observable changes in the climate and environment in general. Each year global carbon dioxide emissions hit a new record, and it is reported that 0.5\% of total US greenhouse gas emissions are attributed to data centres as of 2021 (\cite{Siddik2021TheEF}). 
The release of ChatGPT in late 2022 sparked social interest in Large Language Models (LLMs), the new generation of Language Models with a large number of parameters and trained on massive amounts of data. Currently, numerous companies are releasing products featuring various LLMs, with many more models in development and awaiting release.
Deep Learning research is a competitive field, with only models that reach top performance attracting attention and being utilized. Hence, achieving better accuracy and results is often the first priority, while the model's efficiency and the environmental impact of the study are neglected. However, LLMs demand substantial computational resources and are very costly to train, both financially and environmentally. It becomes essential to raise awareness and promote conscious decisions about algorithmic and hardware choices. 
Providing information on training time, the approximate carbon dioxide emissions and power consumption would assist future studies in making necessary adjustments and determining the compatibility of available computational resources with model requirements.
In this study, we infused T5 LLM with external knowledge and fine-tuned the model for Question-Answering task. Furthermore, we calculated and reported the approximate environmental impact for both steps.
The findings demonstrate that the smaller models may not always be sustainable options, and increased training does not always imply better performance. The most optimal outcome is achieved by carefully considering both performance and efficiency factors. 

\end{abstract}
\vspace{10pt}
\hypertarget{Keywords}{}
\label{Keywords}
\newline
\hypertarget{Introduction}{}
\section{Introduction}
\label{Introduction}


\begin{table}[t]
\begin{center}
\begin{tabular}{ c  c  c  c }
\hline
\textbf{Model} & \textbf{Parameters}  &  \multicolumn{1}{c}{\begin{tabular}[c]{@{}c@{}}\textbf{Estimated}\\ \textbf{emissions} \end{tabular}} & \multicolumn{1}{c}{\begin{tabular}[c]{@{}c@{}}\textbf{Equivalence in} \\ \textbf{\# of flights}\end{tabular}} \\ 
\hline
BERT & 110M & 1.59 & 1.9 \\
HuggingFace, BLOOM & 176B & 25 & 30 \\
META, OPT & 175B & 75 &  90 \\
DeepMind, Gopher & 280B & 380 &  456 \\
Open AI, GPT3 & 175B & 500 & 600 \\
\hline
\end{tabular}
\end{center}
\caption[LLMs environmental impact]{The first column shows state-of-the-art LLMs, along with the corresponding number of parameters in the second column and emissions produced during training in metric tons $CO_2 eq$ in the third column. The last column represents the equivalence in the number of round flights between London and New York.}
\label{t:llm_emissions}
\end{table}

With the growing problem of climate change, LLMs can potentially accelerate that process by contributing to greenhouse gas emissions. LLMs with billions of parameters may require several weeks of training time, and this duration is expected to increase further with the emergence of new models (\cite{scao2022bloom, radford2019language, brown2020language}). \ref{t:llm_emissions} demonstrates the most recent LLMs released by famous research labs, the number of parameters of each model, the estimated emissions in net metric tons $CO_2 eq$ and the equivalence in flights. The amount of produced emissions doubles if taken into consideration the manufacturing of computers. Considering the computational expenses involved, it is only essential to prevent executing identical experiments and adopt a sustainability mindset in research endeavours. This means that researchers have to report not only performance but also training time, energy consumption, pre-training and fine-tuning requirements, and any other metrics that demonstrate the model's efficiency. Reporting training time and energy consumption can help to identify resource-intensive approaches to avoid or optimize them later. Carbon dioxide equivalence helps to assess the environmental impact of the research holistically. Ultimately, understanding and minimizing resource consumption and reducing carbon emissions promote the development of more sustainable practices and making informed decisions towards effective and efficient solutions.


In this study, we focus on Commonsense Reasoning and Question-Answering NLP tasks. Commonsense is a set of implicit pre-knowledge about the everyday world. For example, it is common knowledge that a refrigerator can be found in the kitchen and that summer comes after spring. Commonsense reasoning requires human experience, together with social, physical, temporal and spatial information of everyday life. Learning and using implicit knowledge for humans is an easy everyday task, which makes their language concise yet precise. However, machines do not possess common sense and are not able to learn such knowledge by interacting with the environment. That makes the Commonsense Question Answering (CSQA) task one of the major goals in the Artificial Intelligence (AI) community.


A way to teach models common sense and reasoning is through training them on a commonsense data. The study conducted by \cite{lal2021tellmewhy} introduced a new dataset for CSQA and fine-tuned three LLMs to showcase the TellMeWhy dataset. The authors fine-tuned and tested the performances of T5 (\cite{raffel2020exploring}), GPT 2 (\cite{radford2019language}) and UnifiedQA (\cite{khashabi2020unifiedqa}). To assess the effect of data size, model parameters and points mentioned above, our study similarly explores the T5 model and fine-tunes it on the TellMeWhy dataset. 

In this study, we focus on two aspects: 1) how long does it take to train the T5 model and what environmental impact it has; 2) how does knowledge infusion from Knowledge Graphs (KG) influence the T5 model's ability to perform on CSQA task. Both goals are supposed to be achieved by injecting the commonsense knowledge from KG and fine-tuning the model on the CSQA dataset.

\hypertarget{Related work}{}
\section{Related work}
\label{Related_work}

Many advocate making efficiency reports a routine practice in deep learning research. Yet, when diving deeper into the problem, it is clear that part of the reason why very few researchers report efficiency results is because of the absence of a standard of measurement. There are numerous metrics available to assess the quality of the model, and often times improved performance means a better prediction ability. Some even argue that modern AI does not actually learn and is just a result of utilizing massive amounts of data and large computation power. Although sustainability in AI is still in its infancy, there are already great studies being held to bring awareness to the research community. In this section, we mention works that have been held to quantify and measure the carbon footprint of LLMs. By the end of the section, we will also briefly mention studies in knowledge infusion, which is also part of our study.

Multiple studies have already focused on energy consumption and carbon emissions accounting; some even propose methods for mitigating the problem.
Prioritizing the model's efficiency over performance is becoming more relevant as more powerful machines are being developed. Several factors contribute to the increase in training time directly or indirectly, the development of more robust and powerful hardware, more complex machine learning algorithms and approaches, data growth, and social demand.

The study of \citeauthor{strubell2019} draws attention to the potentially hazardous impacts of training large models on our environment and proposes solutions to mitigate the problem. As an example, they trained a few state-of-the-art LLMs and put them into perspective by quantifying carbon emissions produced during training. Later they compared the results with the emissions produced during a flight and cloud computing prices.
The work has concluded that training BERT emits roughly the same amount of carbon into the atmosphere as a trans-American flight. 

The work of \citeauthor{wu2022sustainable} goes beyond measuring carbon emissions during training. The study also includes model development and inference phases. The authors encourage not only to look at the training phase but to consider the machine learning pipeline end-to-end, starting from data collection until inference. They examine the ML development cycle across the industry scale. Operational and manufacturing carbon footprint is also taken into account, by the end of the study the authors discuss how hardware choices and optimization techniques can help to reduce the carbon footprint of an AI system.

Work conducted by \citeauthor{patterson2021carbon} proved that most of the companies and research groups try to avoid pre-training and prefer executing fine-tuning and inference stages. The study suggests that such stages are as important as pre-training and should not be neglected when it comes to carbon footprint accounting. The study proved that the inference can produce a significant amount of emissions as well.


So far, we have looked into measures for energy consumption and studies conducted on green AI. Now we will inspect KG-infusion methods, as it is a promising approach for carbon footprint reduction by enabling hybrid or neuro-symbolic AI.
According to \citeauthor{bauer2018commonsense} providing LLMs with external knowledge enhances its ability to reason on a downstream task, i.e. QA, summarization, etc. Knowledge infusion enriches the model's vocabulary and allows it to "think out of the box". Some works have already attempted to incorporate commonsense knowledge into the BERT model to enhance reading comprehension (\citeauthor{yang2019enhancing}), and relation classification (\citeauthor{zhang2019ernie}).

A recently conducted study by \citeauthor{lal2022using} extends their previous work on Commonsense QA. The authors utilize COMET KG as an external knowledge source and inject the knowledge into the LLM. As a result, they observed an increase in performance. However, none of the studies measure and report the environmental impact of their work.


\hypertarget{Experimental Setup}{}
\section{Experimental Setup}
\label{Experimental_setup}

\hypertarget{Dataset}{}
\subsection{Dataset}
\label{Dataset}

\subsubsection{Knowledge infusion}

\begin{figure}[t]
\includegraphics[width=\linewidth]{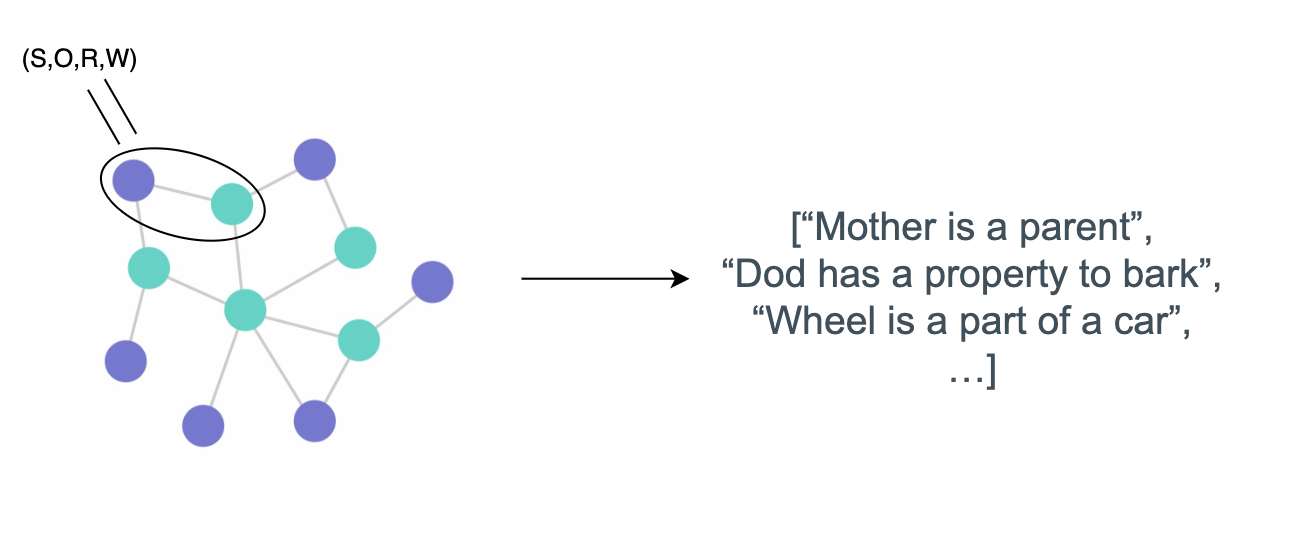}
\caption{Transforming KG into a natural language sentence.}
\label{fig:kg_to_sent}
\end{figure}

 As external knowledge for our LLM, we combined ConceptNet (\citeauthor{speer2017conceptnet}) and ATOMIC (\citeauthor{maarten2018atomic}), which are both large-scale Knowledge Graphs containing information about events in everyday life. Both KGs have to be pre-processed prior to being fed into the LLM. In the scope of this study, we pre-processed and verbalised only ConceptNet KG and combine it with already pre-processed ATOMIC KG \citeauthor{guan2020knowledge}. 

 ConceptNet is constructed of multiple triplets (Subject, Relation, Object) with corresponding relation weight. We start by iterating over subjects and sorting them based on their relationship weight. Then we select the top 100 triplets with respect to relation weight and transform them into sentences using simple verbalization templates (\citeauthor{levy-etal-2017-zero}), see Figure~\ref{fig:kg_to_sent}. The combined pre-processed dataset contains 1,174,267 sentences in the train set and 66,856 in the validation set. The dataset contains physical, spatial, social, and temporal aspects of daily life.

\subsubsection{Fine-tuning}
Following \citeauthor{lal2021tellmewhy} example, models are fine-tuned on the TellMeWhy corpus, the largest Commonsense QA dataset. It incorporates various stories, 30K open-ended questions, and free-form answers. The provided short narratives describe why characters performed certain actions. The answers can be explicit and found in the narrative, as well as implicit, answers that require external knowledge, some intuitive knowledge about the world.



\hypertarget{Methods}{}
\subsection{Methods}
\label{Methods}

\begin{figure}[t]
\includegraphics[width=\linewidth]{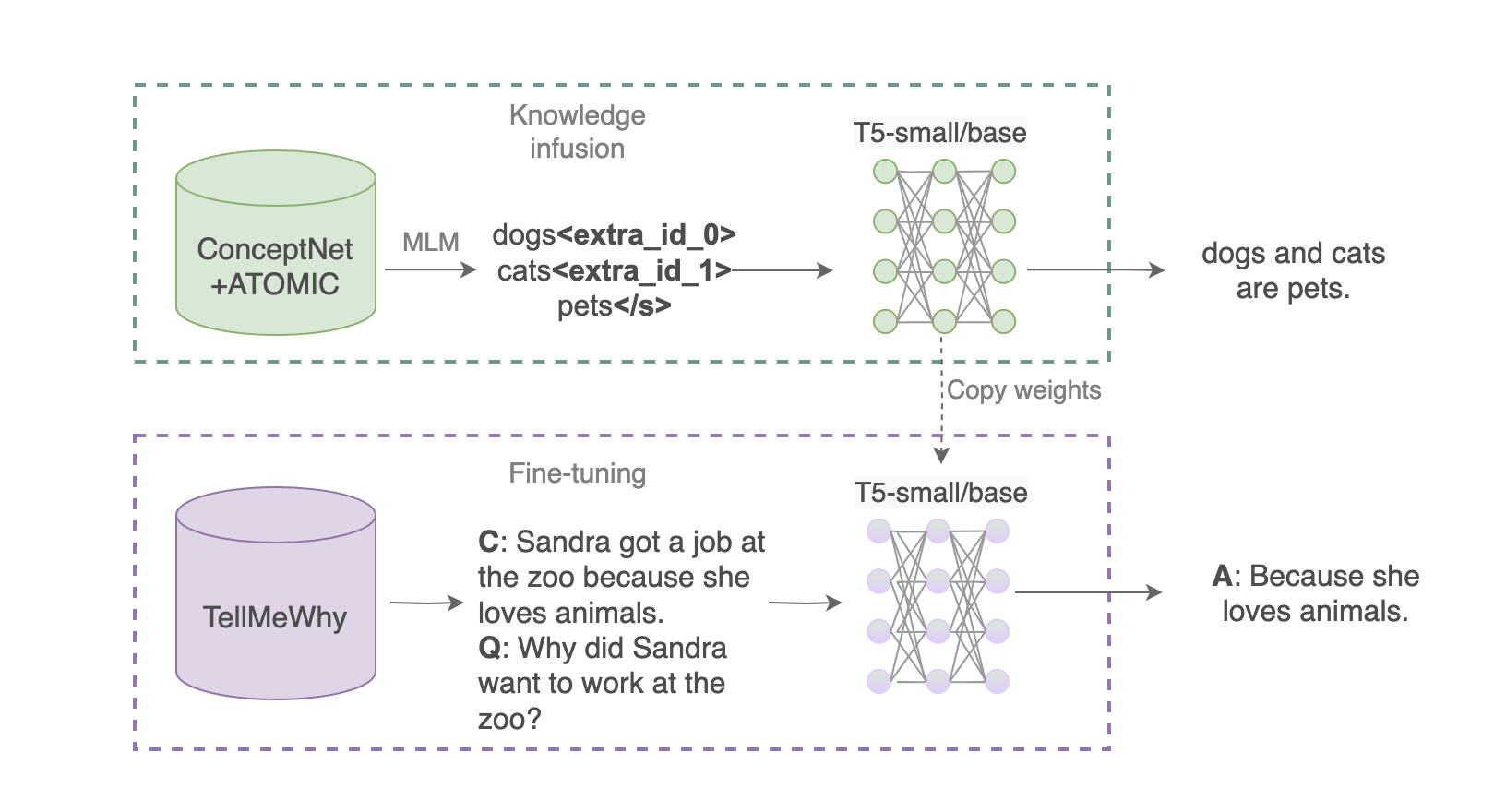}
\caption{Study workflow. First step: Infusing T5-small/-base with knowledge from Knowledge Graphs. Second step: Fine-tuning model with injected knowledge for QA task.}
\label{fig:flowchart}
\end{figure}

As mentioned before, injecting commonsense knowledge from KG beforehand should prepare a solid base for later fine-tuning. Followed up with training on a Question-Answering task-specific TellMeWhy dataset, namely commonsense QA, might result in better rationalization and reasoning abilities.

Originally, T5 was pre-trained on the large unlabelled corpus, Colossal Clean Crawled Corpus (C4) corpus (\citeauthor{raffel2020exploring}) cleaned information from the web. The web-crawled data consists of over 300M values of sentences of various topics; hence, further training the model on commonsense knowledge data might strengthen T5's ability to form constructive sequences and generate better reasoning.
Step 1 is the continuation of T5's original unsupervised pre-training on the Masked Language Modeling task (MLM). The words in encoded sentences are masked with a 15\% probability, together with the reversed masks as labels they are fed into the network. We maintained input and output in the same fashion as the original pre-training. Due to the size of this dataset, the knowledge infusion step runs only for one epoch, more extended training seems to lead to overfitting.

Once the knowledge infusion is completed, the model is fine-tuned for the QA task. Encoded context and question serve as input to produce the predicted answer, which is then compared to the reference answer. This step allows the model to only concentrate on a specific NLP task.

We maintained the same parameters for the fine-tuning phase as \citeauthor{lal2021tellmewhy}. We set the maximum number of epochs to 50, with a learning rate of 5-e5 and a batch size of 16. Our experiments also run until the validation loss does not improve for three iterations. However, we set the maximum source length to 255.

\hypertarget{Models}{}
\subsection{Models}
\label{Models}
In the scope of this study, we utilized  T5-small and T5-base models (\citeauthor{raffel2020exploring}). T5 is a transformer-based model that can be used for multiple NLP tasks without making any architectural changes in contrast to other language models, due to the unified text-to-text format. Such architecture enables the application of transfer learning techniques to reduce the training cost. Both, the input and the output, are string types. Task specifications are added in the beginning and separated by the colon from the input.

The T5-small version has 60 million, and T5-base has 220 million parameters. T5-large with 11B parameters was too computationally expensive; hence it was not utilized in this study.

\begin{table*}[!h]
\hypertarget{fullwidth}{}
\begin{center}
\begin{tabular}{ c c c c c c}
\hline
\textbf{Experiment} & \multicolumn{1}{c}{\begin{tabular}[c]{@{}c@{}} \textbf{Convergence}\\ \textbf{epoch} \end{tabular}} & \textbf{BLEU} $\uparrow$ & \textbf{RG-L F1} $\uparrow$\tnote{1} & \textbf{BLEURT} $\downarrow$\tnote{1} & \textbf{BERTscore} $\uparrow$\tnote{1} \\
\hline
\multicolumn{6}{x{0.64\textwidth}}{Full Test Set} \\
\hline
T5s FT & 12 & 21.93 & 0.25 & -0.412 & 0.501 \\
T5s IK+FT & 13 & 22.94 & 0.25 & -0.374 & 0.513 \\
T5b FT & 6 & 24.43 & 0.26 & -0.359 & \textbf{0.535} \\
T5b IK+FT & 6 & \textbf{24.57} & \textbf{0.26} & -0.3514 & 0.5338  \\
\hline
\citeauthor{lal2021tellmewhy} T5-base & 30-50 & 24.53 & 0.24 & \textbf{-0.28} & 0.48  \\
\hline
\multicolumn{6}{x{0.64\textwidth}}{Implicit-Answer Questions} \\
\hline
T5s FT & 12 & 15.2 & 0.19 & -0.618 & 0.429 \\
T5s IK+FT & 13 & 15.32 & 0.19 & -0.589 & 0.431 \\
T5b FT & 6 & \textbf{16.92} & 0.2 & -0.58 & \textbf{0.452} \\
T5b IK+FT & 6 & 16.53 & \textbf{0.2} & -0.577 & 0.4457 \\
\hline
\citeauthor{lal2021tellmewhy} T5-base & 30-50 & 16.31 & 0.17 & \textbf{-0.51} & 0.34  \\
\hline
\end{tabular}
\end{center}
\caption{Performance of models on the full test set and implicit answer questions in the test set using automatic evaluation provided by \citeauthor{lal2021tellmewhy}.}
\label{t:fullwidthTraining}
\end{table*}

\hypertarget{Results}{}
\section{Results}
\label{Results}

In the scope of our study, we conducted 6 experiments, which will be further referred to as follows:

\begin{enumerate}
    \item \textit{T5s IK}: T5-small with injected knowledge from KG
    \item \textit{T5b IK}: T5-base with injected knowledge from KG
    \item \textit{T5s FT}: T5-small fine-tuned for QA task
    \item \textit{T5b FT}: T5-base fine-tuned for QA task
    \item \textit{T5s IK+FT}: T5-small with injected knowledge from KG and fine-tuned for QA task
    \item \textit{T5b IK+FT}: T5-base with injected knowledge from KG and fine-tuned for QA task
\end{enumerate}

All experiments were performed on 2 NVIDIA RTX A5000 GPU blocks with 24GB memory each. The implementation is available on GitHub\footnote{\url{https://github.com/aidausmanova/commonsense_qa}}.

\hypertarget{Performance Metrics}{}
\subsection{Performance Metrics}
\label{Performance_evaluation}
To evaluate the model's performance, we utilized the same metrics as \citeauthor{lal2021tellmewhy}.
BLEURT (\citeauthor{sellam2020bleurt}) and BLEU (\citeauthor{papineni2002bleu}) scores are both learned evaluation metrics for natural text generation based on BERT. Being trained on WMT human annotations for the machine translation task, they correlate well with human judgments. The scores are generated  based on the precision of tokens of a candidate sentence to the reference.
While BertScore (\citeauthor{zhang2019bertscore}) uses only pre-trained contextual embeddings from BERT and matches words between two sentences by cosine similarity.

We also measured cosine similarity between the generated and the target answers and analysed the number of unique words presented in answer vocabulary that does not exist in context vocabulary.

\hypertarget{Performance analysis}{}
\subsection{Performance analysis}
\label{Performance_analysis}

Table~\ref{t:fullwidthTraining} presents the model performance in various setups, the automatic evaluation provided on the official TellMeWhy GitHub repository\footnote{\url{https://github.com/StonyBrookNLP/tellmewhy}}. Based on the results, we cannot prove that infusing T5 with commonsense knowledge from ConceptNet and ATOMIC influences the model’s ability to reason. This could be due to the large size of the C4 corpus, and, thus, KGs ConceptNet, and ATOMIC failing to provide enough knowledge to teach the network. However, we can conclude that the T5 model is inherently bad at commonsense reasoning, due to the type of data it has been pre-trained on. 

Yet, there is an observable difference in results between \textit{T5s FT} and \textit{T5s IK+FT} across most of the metrics. \textit{T5s IK+FT} performed slightly better than \textit{T5s FT}. The difference in the BLEU score is 1.01 and in BertScore is 2.4\%, both are noticeable differences, considering the evaluation is for similar models and on the same dataset. We assume that due to the smaller size of the T5-small model, the significance of commonsense knowledge from ConcentNet and ATOMIC was more prominent. Compared to T5-base, T5-small seems to gain more from the knowledge infusion step. While for T5-base, ConceptNet and ATOMIC KGs are too small to make a visible difference. 

The BLEURT score demonstrates that for all experiments, there exists a negative correlation between predicted and reference answers. The BLEURT and BLEU scores were specifically designed to assess the quality of the machine translation; this could explain the insignificance of the results. However, since BertScore only uses BERT embeddings and calculates the cosine similarity between two sentences, we observe a higher correlation between the predicted and goal answers.

Similarly to \citeauthor{lal2021tellmewhy}, models perform best when the answer is explicitly given in the context. We observed a slight performance increase for \textit{T5b IK+FT} compared to the \textit{T5 base} results of \citeauthor{lal2021tellmewhy}. We anticipate that the ROUGE F-1 and BertScores scores are higher in our experiments, compared to that of \citeauthor{lal2021tellmewhy}, because we set \textit{max\_len\_seq} to 200, as this was the size of the longest token in our case. However, we still came to the conclusion that the most influence comes from fine-tuning step, but it seems like knowledge infusion makes some difference for smaller models.

It is worth noting that the comparable results in our experiments were achieved with fewer epochs. \citeauthor{lal2021tellmewhy} suggests that T5-base reaches the best performance between epochs 30 and 50. However, we could see that longer training does not add much to the performance and adding EarlyStopping is necessary to prevent not only overfitting but also resource over usage.

The semantic similarity between the answers increases as the training time and size of a model also increase, but the difference is not significant. Surprisingly, infusing models with commonsense knowledge and fine-tuning on QA resulted in the model using more TellMeWhy context vocabulary rather than the model that was just fine-tuned on QA.


\hypertarget{Efficiency Metrics}{}
\subsection{Efficiency Metrics}
\label{Efficiency_evaluation}

Some studies provide great solutions to facilitate carbon emissions and energy consumption calculation. \citeauthor{anthony2020carbontracker} developed a library that accesses information about hardware and calculates the estimates after the first epoch. Their Carbontracker gives information about approximate carbon emissions in grams, energy consumption (KW/h), and an equivalent number of kilometres the car would have driven producing the same amount of emissions. Alternatively, \citeauthor{lacoste2019quantifying} developed a tool that can be used after executing experiments. By providing training time, location, and hardware type, you can estimate produced $CO_2$ emissions and also how much would have been emitted if the experiment was held in a different datacenter.

In our study, we embedded Carbontracker (\citeauthor{anthony2020carbontracker}) into the training loop, which approximates carbon emissions and power usage for the whole training after 1 epoch. The following formula \ref{pt} is used to calculate the power usage of an experiment $p_t$. The average GPU power draw $p_g$ is usually obtained by querying the NVIDIA System Management Interface throughout the run. The value is then multiplied by the number of GPUs $g$ and the Power Usage Effectiveness Coefficient (PUE) (1.55 for Germany\footnote{\url{https://www.statista.com/statistics/1229367/data-center-average-annual-pue-worldwide/}}).

\begin{equation}
\label{pt}
    p_t = \frac{1.55*t*g*p_g}{1000}
\end{equation}

The number of emissions and power consumption depends on the data center location and the local power grid it is connected to. The same experiments executed in two different locations may have different environmental impacts. As of 2022, the power sector emissions in Germany were approximately 380 grams of carbon dioxide produced per kilowatt-hours ($gCO_2/KWh)$ for generated electricity\footnote{\url{https://www.nowtricity.com/country/germany/}}. To get the carbon emissions equivalence estimation (in kg per kilowatt-hour), emissions per hour are multiplied by the experiment's power usage, as shown in Formula~\ref{co2e}. 

\begin{equation}
\label{co2e}
    CO_2e = 0.380*p_t
\end{equation}

We report the overall time it took to execute one experiment, the energy use, carbon dioxide emissions equivalence, and the equivalence in travel by car, in Table \ref{t:efficiency_results}.

\hypertarget{Efficiency analysis}{}
\subsection{Efficiency analysis}
\label{Efficiency_analysis}

\begin{table}[t]
\begin{center}
\begin{tabular}{ c c c c c}
\hline
\textbf{Experiment} & \multicolumn{1}{c}{\begin{tabular}[c]{@{}c@{}}\textbf{Overall}\\ \textbf{time (hr)}\end{tabular}}  & \multicolumn{1}{c}{\begin{tabular}[c]{@{}c@{}} \textbf{Energy}\\ \textbf{use (KWh)} \end{tabular}} & \multicolumn{1}{c}{\begin{tabular}[c]{@{}c@{}}$\mathbf{CO_{2}eq.}$ \\ \textbf{(kg)}\end{tabular}}  & \multicolumn{1}{c}{\begin{tabular}[c]{@{}c@{}}\textbf{Travel by}\\ \textbf{car (km)}\end{tabular}} \\
\hline
\multicolumn{5}{x{0.4\textwidth}}{Step 1 (Pre-training)} \\
\hline
T5s IK & 4.438 & 3.53 & 1.04 & 8.62 \\
T5b IK & 13.969 & 10.72 & 3.15 & 26.18 \\
\hline
\multicolumn{5}{x{0.4\textwidth}}{Step 2 (Fine-tuning)} \\
\hline
T5s FT & 1.981 & 1.52 & 0.45 & 3.72 \\
T5s IK+FT & 6.612 & 5.19 & 1.53 & 12.68 \\
T5b FT & 3.793 & 2.74 & 0.81 & 6.7 \\
T5b IK+FT & 17.718 & 13.42 & 3.95 & 32.80 \\
\hline
\end{tabular}
\end{center}
\caption{Energy and emissions calculated by Carbontracker (\citeauthor{anthony2020carbontracker}).}
\label{t:efficiency_results}
\end{table}

Table~\ref{t:efficiency_results} presents the training time of each experiment and corresponding efficiency metrics calculated by Carbontracker. Since we set an early stopping during the fine-tuning step, none of the experiments reached the maximum number of epochs. \textit{T5s FT} ran until 12, while \textit{T5s IK+FT} until 13, both \textit{T5b FT} and \textit{T5b IK+FT} stopped after 6 epochs. This fact demonstrates the importance of early stopping in research to prevent unnecessary resource waste and energy consumption when the models do not need long training.

Clearly, the pre-training phase required a much longer training time, the difference between minor and base variants is also significant, with T5-base requiring 3 times more hours to complete 1 epoch. Having a look at the fine-tuning stage, we can see that the difference between \textit{T5s FT} and \textit{T5b FT} is around 2 hours, but \textit{T5b FT} outperforms the former. In our case, pre-training and fine-tuning T5-small did not give a desirable performance, hence T5-base is preferred even with a longer training time. 

Looking at \textit{T5b FT} and \textit{T5b IK+FT}, we noticed that the latter outperforms the first one only by a mere percentage based on BLEU, F1 and BLEURT scores. On the contrary, BERTscore for \textit{T5b FT} has been consistently higher than that for \textit{T5b IK+FT}. 
\hypertarget{Conclusion_and_Discussion}{}
\section{Conclusion}
\label{Conclusion_and_Discussion}
Numerous factors could have influenced the outcome of our study. We assume that among these factors, the nature of the data that we infused into our model influenced the most. While sorting the KG based on relation weight and extracting top N triples seems like a straightforward approach, it yields suboptimal results. The main limitation lies in the lack of diversity within the dataset, with many sentences being semantically close and having limited number of relationship types. 


\begin{itemize}
    \item When looking for a balance between performance and efficiency, \textit{T5b FT} seems like a more reasonable choice.
    \item We need more sophisticated approaches to linearize KGs in a meaningful way
\end{itemize}

Our study showed that it is important to consider a model not solely based on one parameter. Focusing only on performance could lead to uncontrollable energy waste, while trying to reduce energy consumption too much can lead to a weak model that is less sustainable in the long run. The balance between the two is the key to the most optimal solution.

Tracking carbon footprint at every stage of the study is an extremely challenging task and has much more room for improvement regarding the report standards. To get the full picture, one might also need to know how much it takes to build hardware, transport them to the data center, as well as consider the lighting in the room, etc. Nevertheless, it is important to be aware of the factors that impact the quantity of carbon emissions produced by research. As we have seen, stages like fine-tuning can also produce an observable amount of emissions. Such a step towards a positive change can also greatly help follow-up studies in the field.

\section*{Limitations}
Our study includes several limitations that couldn't been addressed in this study and could be an idea for future work.
Firstly, the Knowledge Infusion part of our study did not yield desirable results due to the poor KG linearization strategy. This stage also took the most time to be executed and consumed the most computational power.
Secondly, Due to server limitations, we couldn't perform any experiments on T5-large model, which restricts us from making bolder statements on LLM performance on the CSQA task. 
In this work, we wanted to draw attention to the importance of considering efficiency results together with the performance results of the study.

\section*{Ethics Statement}
Our research focuses on accounting and reporting the environmental impact of LLMs. Such studies raise concerns about transparency and accountability of Deep Learning approaches. It is crucial that the processes and algorithms used in any study are transparent and open to scrutiny. We commit to making our methods and data publicly available for review and validation by the broader community.

While the benefits of this research field are clear, it is essential to acknowledge and address potential ethical considerations. Calculating the exact amount of emitted carbon into the atmosphere presents a challenging task that requires acquiring server production and transportation information, as well as considering local energy grid and its fuel type. Furthermore, we should also scrutinise the Deep Learning model exploitation and life-cycle periods to get a clearer picture of its environmental impact. Hence, this field of study still requires extensive research with its potential positive impact on the research community.

\textbf{Acknowledgements}

This work has been partially supported by grants for the DFG project NFDI4DataScience project (DFG project no. 460234259) and by the Federal Ministry for Economics and Climate Action in the project CoyPu (project number 01MK21007G).

\printbibliography

@inproceedings{lal2021tellmewhy,
  author       = {Yash Kumar Lal and
                  Nathanael Chambers and
                  Raymond J. Mooney and
                  Niranjan Balasubramanian},
  editor       = {Chengqing Zong and
                  Fei Xia and
                  Wenjie Li and
                  Roberto Navigli},
  title        = {TellMeWhy: {A} Dataset for Answering Why-Questions in Narratives},
  booktitle    = {Findings of the Association for Computational Linguistics: {ACL/IJCNLP}
                  2021, Online Event, August 1-6, 2021},
  series       = {Findings of {ACL}},
  volume       = {{ACL/IJCNLP} 2021},
  pages        = {596--610},
  publisher    = {Association for Computational Linguistics},
  year         = {2021},
  url          = {https://doi.org/10.18653/v1/2021.findings-acl.53},
  doi          = {10.18653/v1/2021.findings-acl.53},
  bibsource    = {dblp computer science bibliography, https://dblp.org}
}

@inproceedings{lal2022using,
  author       = {Yash Kumar Lal and
                  Niket Tandon and
                  Tanvi Aggarwal and
                  Horace Liu and
                  Nathanael Chambers and
                  Raymond J. Mooney and
                  Niranjan Balasubramanian},
  editor       = {Yoav Goldberg and
                  Zornitsa Kozareva and
                  Yue Zhang},
  title        = {Using Commonsense Knowledge to Answer Why-Questions},
  booktitle    = {Proceedings of the 2022 Conference on Empirical Methods in Natural
                  Language Processing, {EMNLP} 2022, Abu Dhabi, United Arab Emirates,
                  December 7-11, 2022},
  pages        = {1204--1219},
  publisher    = {Association for Computational Linguistics},
  year         = {2022},
  url          = {https://aclanthology.org/2022.emnlp-main.79},
  bibsource    = {dblp computer science bibliography, https://dblp.org}
}

@inproceedings{speer2017conceptnet,
  author       = {Robyn Speer and
                  Joshua Chin and
                  Catherine Havasi},
  editor       = {Satinder Singh and
                  Shaul Markovitch},
  title        = {ConceptNet 5.5: An Open Multilingual Graph of General Knowledge},
  booktitle    = {Proceedings of the Thirty-First {AAAI} Conference on Artificial Intelligence,
                  February 4-9, 2017, San Francisco, California, {USA}},
  pages        = {4444--4451},
  publisher    = {{AAAI} Press},
  year         = {2017},
  url          = {http://aaai.org/ocs/index.php/AAAI/AAAI17/paper/view/14972},
  bibsource    = {dblp computer science bibliography, https://dblp.org}
}

@inproceedings{maarten2018atomic,
  author       = {Maarten Sap and
                  Ronan Le Bras and
                  Emily Allaway and
                  Chandra Bhagavatula and
                  Nicholas Lourie and
                  Hannah Rashkin and
                  Brendan Roof and
                  Noah A. Smith and
                  Yejin Choi},
  title        = {{ATOMIC:} An Atlas of Machine Commonsense for If-Then Reasoning},
  booktitle    = {The Thirty-Third {AAAI} Conference on Artificial Intelligence, {AAAI}
                  2019, The Thirty-First Innovative Applications of Artificial Intelligence
                  Conference, {IAAI} 2019, The Ninth {AAAI} Symposium on Educational
                  Advances in Artificial Intelligence, {EAAI} 2019, Honolulu, Hawaii,
                  USA, January 27 - February 1, 2019},
  pages        = {3027--3035},
  publisher    = {{AAAI} Press},
  year         = {2019},
  url          = {https://doi.org/10.1609/aaai.v33i01.33013027},
  doi          = {10.1609/aaai.v33i01.33013027},
  bibsource    = {dblp computer science bibliography, https://dblp.org}
}

@article{guan2020knowledge,
  author       = {Jian Guan and
                  Fei Huang and
                  Minlie Huang and
                  Zhihao Zhao and
                  Xiaoyan Zhu},
  title        = {A Knowledge-Enhanced Pretraining Model for Commonsense Story Generation},
  journal      = {Trans. Assoc. Comput. Linguistics},
  volume       = {8},
  pages        = {93--108},
  year         = {2020},
  url          = {https://doi.org/10.1162/tacl\_a\_00302},
  doi          = {10.1162/tacl\_a\_00302},
  bibsource    = {dblp computer science bibliography, https://dblp.org}
}

@article{levy-etal-2017-zero,
  author       = {Omer Levy and
                  Minjoon Seo and
                  Eunsol Choi and
                  Luke Zettlemoyer},
  title        = {Zero-Shot Relation Extraction via Reading Comprehension},
  journal      = {CoRR},
  volume       = {abs/1706.04115},
  year         = {2017},
  url          = {http://arxiv.org/abs/1706.04115},
  eprinttype    = {arXiv},
  eprint       = {1706.04115},
  bibsource    = {dblp computer science bibliography, https://dblp.org}
}

@inproceedings{bauer2018commonsense,
  author       = {Lisa Bauer and
                  Yicheng Wang and
                  Mohit Bansal},
  editor       = {Ellen Riloff and
                  David Chiang and
                  Julia Hockenmaier and
                  Jun'ichi Tsujii},
  title        = {Commonsense for Generative Multi-Hop Question Answering Tasks},
  booktitle    = {Proceedings of the 2018 Conference on Empirical Methods in Natural
                  Language Processing, Brussels, Belgium, October 31 - November 4, 2018},
  pages        = {4220--4230},
  publisher    = {Association for Computational Linguistics},
  year         = {2018},
  url          = {https://doi.org/10.18653/v1/d18-1454},
  doi          = {10.18653/v1/d18-1454},
  bibsource    = {dblp computer science bibliography, https://dblp.org}
}

@inproceedings{yang2019enhancing,
  author       = {An Yang and
                  Quan Wang and
                  Jing Liu and
                  Kai Liu and
                  Yajuan Lyu and
                  Hua Wu and
                  Qiaoqiao She and
                  Sujian Li},
  editor       = {Anna Korhonen and
                  David R. Traum and
                  Llu{\'{\i}}s M{\`{a}}rquez},
  title        = {Enhancing Pre-Trained Language Representations with Rich Knowledge
                  for Machine Reading Comprehension},
  booktitle    = {Proceedings of the 57th Conference of the Association for Computational
                  Linguistics, {ACL} 2019, Florence, Italy, July 28- August 2, 2019,
                  Volume 1: Long Papers},
  pages        = {2346--2357},
  publisher    = {Association for Computational Linguistics},
  year         = {2019},
  url          = {https://doi.org/10.18653/v1/p19-1226},
  doi          = {10.18653/v1/p19-1226},
  bibsource    = {dblp computer science bibliography, https://dblp.org}
}

@inproceedings{zhang2019ernie,
  author       = {Zhengyan Zhang and
                  Xu Han and
                  Zhiyuan Liu and
                  Xin Jiang and
                  Maosong Sun and
                  Qun Liu},
  editor       = {Anna Korhonen and
                  David R. Traum and
                  Llu{\'{\i}}s M{\`{a}}rquez},
  title        = {{ERNIE:} Enhanced Language Representation with Informative Entities},
  booktitle    = {Proceedings of the 57th Conference of the Association for Computational
                  Linguistics, {ACL} 2019, Florence, Italy, July 28- August 2, 2019,
                  Volume 1: Long Papers},
  pages        = {1441--1451},
  publisher    = {Association for Computational Linguistics},
  year         = {2019},
  url          = {https://doi.org/10.18653/v1/p19-1139},
  doi          = {10.18653/v1/p19-1139},
  bibsource    = {dblp computer science bibliography, https://dblp.org}
}

@article{raffel2020exploring,
  author       = {Colin Raffel and
                  Noam Shazeer and
                  Adam Roberts and
                  Katherine Lee and
                  Sharan Narang and
                  Michael Matena and
                  Yanqi Zhou and
                  Wei Li and
                  Peter J. Liu},
  title        = {Exploring the Limits of Transfer Learning with a Unified Text-to-Text
                  Transformer},
  journal      = {J. Mach. Learn. Res.},
  volume       = {21},
  pages        = {140:1--140:67},
  year         = {2020},
  url          = {http://jmlr.org/papers/v21/20-074.html},
  bibsource    = {dblp computer science bibliography, https://dblp.org}
}

@inproceedings{khashabi2020unifiedqa,
  author       = {Daniel Khashabi and
                  Sewon Min and
                  Tushar Khot and
                  Ashish Sabharwal and
                  Oyvind Tafjord and
                  Peter Clark and
                  Hannaneh Hajishirzi},
  editor       = {Trevor Cohn and
                  Yulan He and
                  Yang Liu},
  title        = {UnifiedQA: Crossing Format Boundaries With a Single {QA} System},
  booktitle    = {Findings of the Association for Computational Linguistics: {EMNLP}
                  2020, Online Event, 16-20 November 2020},
  series       = {Findings of {ACL}},
  volume       = {{EMNLP} 2020},
  pages        = {1896--1907},
  publisher    = {Association for Computational Linguistics},
  year         = {2020},
  url          = {https://doi.org/10.18653/v1/2020.findings-emnlp.171},
  doi          = {10.18653/v1/2020.findings-emnlp.171},
  bibsource    = {dblp computer science bibliography, https://dblp.org}
}

@article{scao2022bloom,
  author       = {Teven Le Scao and
                  Angela Fan and
                  Christopher Akiki and
                  Ellie Pavlick and
                  Suzana Ilic and
                  Daniel Hesslow and
                  Roman Castagn{\'{e}} and
                  Alexandra Sasha Luccioni and
                  Fran{\c{c}}ois Yvon and
                  Matthias Gall{\'{e}} and
                  Jonathan Tow and
                  Alexander M. Rush and
                  Stella Biderman and
                  Albert Webson and
                  Pawan Sasanka Ammanamanchi and
                  Thomas Wang and
                  Beno{\^{\i}}t Sagot and
                  Niklas Muennighoff and
                  Albert Villanova del Moral and
                  Olatunji Ruwase and
                  Rachel Bawden and
                  Stas Bekman and
                  Angelina McMillan{-}Major and
                  Iz Beltagy and
                  Huu Nguyen and
                  Lucile Saulnier and
                  Samson Tan and
                  Pedro Ortiz Suarez and
                  Victor Sanh and
                  Hugo Lauren{\c{c}}on and
                  Yacine Jernite and
                  Julien Launay and
                  Margaret Mitchell and
                  Colin Raffel and
                  Aaron Gokaslan and
                  Adi Simhi and
                  Aitor Soroa and
                  Alham Fikri Aji and
                  Amit Alfassy and
                  Anna Rogers and
                  Ariel Kreisberg Nitzav and
                  Canwen Xu and
                  Chenghao Mou and
                  Chris Emezue and
                  Christopher Klamm and
                  Colin Leong and
                  Daniel van Strien and
                  David Ifeoluwa Adelani and
                  et al.},
  title        = {{BLOOM:} {A} 176B-Parameter Open-Access Multilingual Language Model},
  journal      = {CoRR},
  volume       = {abs/2211.05100},
  year         = {2022},
  url          = {https://doi.org/10.48550/arXiv.2211.05100},
  doi          = {10.48550/arXiv.2211.05100},
  eprinttype    = {arXiv},
  eprint       = {2211.05100},
  bibsource    = {dblp computer science bibliography, https://dblp.org}
}

@article{radford2019language,
  title={Language models are unsupervised multitask learners},
  author={Radford, Alec and Wu, Jeffrey and Child, Rewon and Luan, David and Amodei, Dario and Sutskever, Ilya and others},
  journal={OpenAI blog},
  volume={1},
  number={8},
  pages={9},
  year={2019}
}

@inproceedings{brown2020language,
  author       = {Tom B. Brown and
                  Benjamin Mann and
                  Nick Ryder and
                  Melanie Subbiah and
                  Jared Kaplan and
                  Prafulla Dhariwal and
                  Arvind Neelakantan and
                  Pranav Shyam and
                  Girish Sastry and
                  Amanda Askell and
                  Sandhini Agarwal and
                  Ariel Herbert{-}Voss and
                  Gretchen Krueger and
                  Tom Henighan and
                  Rewon Child and
                  Aditya Ramesh and
                  Daniel M. Ziegler and
                  Jeffrey Wu and
                  Clemens Winter and
                  Christopher Hesse and
                  Mark Chen and
                  Eric Sigler and
                  Mateusz Litwin and
                  Scott Gray and
                  Benjamin Chess and
                  Jack Clark and
                  Christopher Berner and
                  Sam McCandlish and
                  Alec Radford and
                  Ilya Sutskever and
                  Dario Amodei},
  editor       = {Hugo Larochelle and
                  Marc'Aurelio Ranzato and
                  Raia Hadsell and
                  Maria{-}Florina Balcan and
                  Hsuan{-}Tien Lin},
  title        = {Language Models are Few-Shot Learners},
  booktitle    = {Advances in Neural Information Processing Systems 33: Annual Conference
                  on Neural Information Processing Systems 2020, NeurIPS 2020, December
                  6-12, 2020, virtual},
  year         = {2020},
  url          = {https://proceedings.neurips.cc/paper/2020/hash/1457c0d6bfcb4967418bfb8ac142f64a-Abstract.html},
  bibsource    = {dblp computer science bibliography, https://dblp.org}
}

@article{anthony2020carbontracker,
  author       = {Lasse F. Wolff Anthony and
                  Benjamin Kanding and
                  Raghavendra Selvan},
  title        = {Carbontracker: Tracking and Predicting the Carbon Footprint of Training
                  Deep Learning Models},
  journal      = {CoRR},
  volume       = {abs/2007.03051},
  year         = {2020},
  url          = {https://arxiv.org/abs/2007.03051},
  eprinttype    = {arXiv},
  eprint       = {2007.03051},
  bibsource    = {dblp computer science bibliography, https://dblp.org}
}

@article{lacoste2019quantifying,
  author       = {Alexandre Lacoste and
                  Alexandra Luccioni and
                  Victor Schmidt and
                  Thomas Dandres},
  title        = {Quantifying the Carbon Emissions of Machine Learning},
  journal      = {CoRR},
  volume       = {abs/1910.09700},
  year         = {2019},
  url          = {http://arxiv.org/abs/1910.09700},
  eprinttype    = {arXiv},
  eprint       = {1910.09700},
  bibsource    = {dblp computer science bibliography, https://dblp.org}
}

@inproceedings{strubell2019,
  author       = {Emma Strubell and
                  Ananya Ganesh and
                  Andrew McCallum},
  editor       = {Anna Korhonen and
                  David R. Traum and
                  Llu{\'{\i}}s M{\`{a}}rquez},
  title        = {Energy and Policy Considerations for Deep Learning in {NLP}},
  booktitle    = {Proceedings of the 57th Conference of the Association for Computational
                  Linguistics, {ACL} 2019, Florence, Italy, July 28- August 2, 2019,
                  Volume 1: Long Papers},
  pages        = {3645--3650},
  publisher    = {Association for Computational Linguistics},
  year         = {2019},
  url          = {https://doi.org/10.18653/v1/p19-1355},
  doi          = {10.18653/v1/p19-1355},
  bibsource    = {dblp computer science bibliography, https://dblp.org}
}

@article{patterson2021carbon,
  author       = {David A. Patterson and
                  Joseph Gonzalez and
                  Quoc V. Le and
                  Chen Liang and
                  Lluis{-}Miquel Munguia and
                  Daniel Rothchild and
                  David R. So and
                  Maud Texier and
                  Jeff Dean},
  title        = {Carbon Emissions and Large Neural Network Training},
  journal      = {CoRR},
  volume       = {abs/2104.10350},
  year         = {2021},
  url          = {https://arxiv.org/abs/2104.10350},
  eprinttype    = {arXiv},
  eprint       = {2104.10350},
  bibsource    = {dblp computer science bibliography, https://dblp.org}
}

@inproceedings{wu2022sustainable,
  author       = {Carole{-}Jean Wu and
                  Ramya Raghavendra and
                  Udit Gupta and
                  Bilge Acun and
                  Newsha Ardalani and
                  Kiwan Maeng and
                  Gloria Chang and
                  Fiona Aga Behram and
                  Jinshi Huang and
                  Charles Bai and
                  Michael Gschwind and
                  Anurag Gupta and
                  Myle Ott and
                  Anastasia Melnikov and
                  Salvatore Candido and
                  David Brooks and
                  Geeta Chauhan and
                  Benjamin Lee and
                  Hsien{-}Hsin S. Lee and
                  Bugra Akyildiz and
                  Maximilian Balandat and
                  Joe Spisak and
                  Ravi Jain and
                  Mike Rabbat and
                  Kim M. Hazelwood},
  editor       = {Diana Marculescu and
                  Yuejie Chi and
                  Carole{-}Jean Wu},
  title        = {Sustainable {AI:} Environmental Implications, Challenges and Opportunities},
  booktitle    = {Proceedings of Machine Learning and Systems 2022, MLSys 2022, Santa
                  Clara, CA, USA, August 29 - September 1, 2022},
  publisher    = {mlsys.org},
  year         = {2022},
  url          = {https://proceedings.mlsys.org/paper/2022/hash/ed3d2c21991e3bef5e069713af9fa6ca-Abstract.html},
  bibsource    = {dblp computer science bibliography, https://dblp.org}
}

@article{Siddik2021TheEF,
  title={The environmental footprint of data centers in the United States},
  author={Md. Abu Bakar Siddik and Arman Shehabi and Landon T. Marston},
  journal={Environmental Research Letters},
  year={2021},
  volume={16}
}

@inproceedings{sellam2020bleurt,
  author       = {Thibault Sellam and
                  Dipanjan Das and
                  Ankur P. Parikh},
  editor       = {Dan Jurafsky and
                  Joyce Chai and
                  Natalie Schluter and
                  Joel R. Tetreault},
  title        = {{BLEURT:} Learning Robust Metrics for Text Generation},
  booktitle    = {Proceedings of the 58th Annual Meeting of the Association for Computational
                  Linguistics, {ACL} 2020, Online, July 5-10, 2020},
  pages        = {7881--7892},
  publisher    = {Association for Computational Linguistics},
  year         = {2020},
  url          = {https://doi.org/10.18653/v1/2020.acl-main.704},
  doi          = {10.18653/v1/2020.acl-main.704},
  bibsource    = {dblp computer science bibliography, https://dblp.org}
}

@inproceedings{papineni2002bleu,
  author       = {Kishore Papineni and
                  Salim Roukos and
                  Todd Ward and
                  Wei{-}Jing Zhu},
  title        = {Bleu: a Method for Automatic Evaluation of Machine Translation},
  booktitle    = {Proceedings of the 40th Annual Meeting of the Association for Computational
                  Linguistics, July 6-12, 2002, Philadelphia, PA, {USA}},
  pages        = {311--318},
  publisher    = {{ACL}},
  year         = {2002},
  url          = {https://aclanthology.org/P02-1040/},
  doi          = {10.3115/1073083.1073135},
  bibsource    = {dblp computer science bibliography, https://dblp.org}
}

@inproceedings{zhang2019bertscore,
  author       = {Tianyi Zhang and
                  Varsha Kishore and
                  Felix Wu and
                  Kilian Q. Weinberger and
                  Yoav Artzi},
  title        = {BERTScore: Evaluating Text Generation with {BERT}},
  booktitle    = {8th International Conference on Learning Representations, {ICLR} 2020,
                  Addis Ababa, Ethiopia, April 26-30, 2020},
  publisher    = {OpenReview.net},
  year         = {2020},
  url          = {https://openreview.net/forum?id=SkeHuCVFDr},
  bibsource    = {dblp computer science bibliography, https://dblp.org}
}
\end{document}